\documentclass{elsarticle}

\usepackage{graphicx}

\usepackage{amsmath}
\usepackage{amssymb}
\usepackage{latexsym}
\newtheorem{thm}{Theorem}
\newtheorem{property}[thm]{Property}

\newtheorem{prop}[thm]{Proposition}

\newtheorem{problem}[thm]{Problem}
\newproof{pf}{Proof}
\newdefinition{defn}[thm]{Definition}
\usepackage{threeparttable}
\usepackage{algorithm}
\usepackage{algorithmic}
\usepackage{subfigure}

\begin{document}

\begin{frontmatter}


\title{Granular association rules on two universes with four measures}

\author{Fan Min}
\ead{minfanphd@163.com}
\address{Deparment of Computer Science, Southwest Petroleum University, Chengdu 610500, China}


\begin{abstract}
Relational association rules reveal patterns hidden in multiple tables.
Existing rules are usually evaluated through two measures, namely support and confidence.
However, these two measures may not be enough to describe the strength of a rule.
In this paper, we introduce granular association rules with four measures to reveal connections between granules in two universes, and propose three algorithms for rule mining.
An example of such rule might be ``40\% men like at least 30\% kinds of alcohol; 45\% customers are men and 6\% products are alcohol."
Here 45\%, 6\%, 40\%, and 30\% are the \emph{source coverage}, the \emph{target coverage}, the \emph{source confidence}, and the \emph{target confidence}, respectively.
With these measures, our rules are semantically richer than existing ones.
Three subtypes of rules are obtained through considering special requirements on the source/target confidence.
Then we define a rule mining problem, and design a sandwich algorithm with different rule checking approaches for different subtypes.
Experiments on two real world datasets show that the approaches dedicated to three subtypes are 2-3 orders of magnitudes faster than the one for the general case.
A forward algorithm and a backward algorithm for one particular subtype can speed up the mining process further.
This work opens a new research trend concerning relational association rule mining, granular computing and cold-start recommendation.
\end{abstract}


\begin{keyword}
Complete match, granule, granular computing, measure, partial match, relational association rule.
\end{keyword}
\end{frontmatter}

  %
  %
  \section{Introduction}\label{section: introduction}
Relational data mining approaches \cite{Berzal2001TBAR,DzeroskiS2003Multi,DzeroskiS2001Relational} look for patterns that involve multiple tables in the database.
Important issues include relational association rule discovery (see, e.g., \cite{AftratiF2012Chains,DehaspeD1999Discovery,DehapseL1998Finding,Goethals2010Mining,Goethals2008Mining,Jensen2000Frequent}), relational decision trees (see, e.g., \cite{BlockeelH1998Top,KramerS1996Structural}), and relational distance-based learning (see, e.g., \cite{EmdeW1996Relational}).
These issues are undoubtedly more general and more challenging than their counterparts on a single data table.
Therefore they become popular in recent years.

People have proposed various types of relational association rules for different applications.
For example, Dehaspe et al. \cite{DehapseL1998Finding} chained binary relations to produce ternary relations, quaternary relations, etc, and then constructed rules from new relations.
Jensen et al. \cite{Jensen2000Frequent} joined a number of primary tables through the central relationship table, and then constructed rules from the new table.
Goethals et al. \cite{Goethals2008Mining} constructed rules from two queries, where one asks for a set of tuples satisfying a certain condition, and the other asks for those tuples satisfying a more specific condition.
Kavurucu et al. \cite{Kavurucu2009ILP} employed relational association rule mining techniques to build a predicative concept learning Inductive Logic Programming (ILP) system.
Goethals et al. \cite{Goethals2010Mining} also constructed rules from frequent itemsets across entities and binary relations, with a key specified such that the occurrences of itemsets are counted in one entity table.

These rules are usually evaluated through two measures, namely \emph{support} and \emph{confidence}, which are well defined for association rules \cite{AgrawalR1993Mining,Luna2013Grammar,SrikantR1996Mining} in a single data table.
Unfortunately, these two measures may not be enough to describe the strength of a relational association rule.
For example, according to \cite{Goethals2010Mining} we may obtain a rule ``75\% female professors teach courses with 10 credits, among 30\% of all courses."
In fact, a professor may teach only one course with 10 credits, or she may teach all courses with 10 credits.
Neither measure distinguishes this kind of difference.

In this paper, we introduce granular association rules with four measures to reveal connections between granules in two universes.
The term ``granular" comes from granular computing \cite{Lin98Granular,YaoVasilakosPedrycz2013Granular,Yao00Granular,Zadeh97Towards,ZhuWang03Reduction}, which is an emerging conceptual and computing paradigm of information processing \cite{BargielaPedrycz02Granular}.
Some people study the granular computing models, such as the partition model \cite{Yao00Granular}, the covering model \cite{ZhuWang03Reduction}.
Some study granular computing approaches, such as rule induction \cite{YaoYao02Induction}, multi-scale feature selection \cite{WuLeung2013optimal}.
Let us consider a many-to-many entity-relationship system with two entities \texttt{customer} and \texttt{product} connected by a relation \texttt{buys}.
``Men," ``young men," and ``Chinese women" are granules of customers.
``Alcohol," ``France alcohol," and ``white stuff" are granules of products.
Examples of granular association rules include ``men like alcohol," ``young men like France alcohol," and ``Chinese women like white stuff."
From the viewpoint of granular computing, these rules are partially ordered.
The first rule is \emph{coarser} than the second one, because ``men" is coarser than ``young men," and ``alcohol" is coarser than ``France alcohol."
The third rule is neither finer nor coarser than the second one.

We propose four measures to evaluate the quality of a granular association rule.
An example of such rule might be ``40\% men like at least 30\% kinds of alcohol; 45\% customers are men and 6\% products are alcohol."
Here 45\%, 6\%, 40\%, and 30\% are the \emph{source coverage}, the \emph{target coverage}, the \emph{source confidence}, and the \emph{target confidence}, respectively.
The \emph{support} measure, which is well defined for other association rules, is redundant since it is equal to the product of the source coverage and the source confidence.
With these four measures, the strength of the rule is well defined.
This is one reason why the new type of rules is semantically richer than most of the existing ones.

Granular association rule is an intersection of relational data mining, granular computing and recommender system.
First, it fills the gap between quantitative association rules and general relational association rules which span across more than two universes (see, e.g., \cite{DehapseL1998Finding,Jensen2000Frequent}) or even the whole database (see, e.g., \cite{Goethals2010Mining,Goethals2008Mining}).
Second, it serves as a substantial application of granular computing, which is currently more a theoretical perspective than a coherent set of methods or principles.
The description of information granules with different attribute-value pairs and different size embodies the essences of granular computing.
Moreover, most existing works of granular computing focus on one universe.
Only a few works of granular computing discuss two universes (see., e.g., \cite{LiZhang08Rough,Lin98Granular,YaoYY04APartition}), let alone respective applications.
Third, it provides a new means to build recommender systems.
Existing cold-start recommendation approaches consider the new user problem or the new item problem, while our approach can be applied to the situation where both user and item are new.

In some cases the source confidence and/or the target confidence might be 100\%, resulting in three subtypes with some properties.
When the source confidence is 100\%, the rule is called a right-hand side partial match one.
When the target confidence is 100\%, the rule is called a left-hand side partial match one.
When both measures are 100\%, the rule is called a complete match one.
In correspondence with these terms, when neither measure is 100\%, the rule is called a partial match one.
We may also view partial match rules as a general case without requirements on the source confidence and the target confidence.

Our objective is to mine all granular association rules satisfying thresholds of four measures.
We design a sandwich rule mining algorithm for this purpose.
With this algorithm, candidate granules are generated in each universe according to the source coverage and target coverage thresholds using existing algorithms such as Apriori \cite{AgrawalR1994Apriori} or FP-growth \cite{HanPeiYin00FP}.
Then candidate rules are generated and checked.
Rules meeting the source confidence and the target confidence thresholds are output.
The rule checking approach for partial match rules is inefficient for other subtypes.
Therefore we design different rule checking approaches for three subtypes to fully take advantage of their characteristics.

We also design two more algorithms to mine complete match rules.
They are called the forward algorithm and the backward algorithm, respectively.
Lower approximation, which is a key concept in rough sets \cite{Pawlak82Rough}, is employed to analyze both algorithms.
Hence granular association rule mining can be viewed as a new application of rough sets.

Experiments are undertaken on the MovieLens \cite{movielens} data set assembled by the GroupLens project \cite{grouplens} and the course selection data from Minnan Normal University.
Some interesting rules are obtained through setting reasonable thresholds of four measures.
The tradeoff between the source confidence and the target confidence of a rule is illustrated.
The efficiencies of different approaches are compared through different settings on four thresholds.
For the sandwich algorithm, rule checking approaches designed for three subtypes are 2-3 orders of magnitude faster than the one for the general case.
Moreover, a forward algorithm and a backward algorithm, which are valid for complete match, can enhance the performance.

The rest of the paper is organized as follows.
Section \ref{section: related-works} reviews three types of classical association rules and four types of relational association rules.
Section \ref{section: granular-rules} defines the data model for granular association rules and three subtypes of rules.
Then Section \ref{section: algorithms} defines the problem and presents a sandwich algorithm for the problem.
A forward algorithm and a backward algorithm are also designed to mine complete match rules.
Experiments on the course selection data are discussed in Section \ref{section: experiments}.
Finally, Section \ref{section: conclusion} presents the concluding remarks and further research directions.

  %
  %
  \section{Related works}\label{section: related-works}
In this section, we review popular association rule mining problems and respective approaches.
We will begin with association rules in a single data table, and then proceed to association rules involving multiple tables.

  %
  %
  \subsection{Association rules}\label{subsection: association-rules}
Association rules on a \emph{single} data table have been well-studied.
They are boolean association rules, quantitative association rules, and multi-level association rules.

  %
  %
  \subsubsection{Boolean association rules}\label{subsubsection: boolean-association-rules}
The concept of association rule was first introduced in \cite{AgrawalR1993Mining} to mine transaction data of a supermarket.
This concept was renamed as \emph{boolean association rule} \cite{SrikantR1996Mining} to distinguish from other types of association rules.
The transaction data, also called the basket data, store items purchased on a per-transaction basis.
An example of such rule is ``30\% of transactions that contain beer also contain diapers; 2\% of all transactions contain both of these items."
Here 30\% and 2\% are the \emph{confidence} and the \emph{support}, respectively of the rule.

From the set point of view, boolean association rules reveal the connection between two disjoint subsets of the same universe.
Let the number of transactions be $n$ and the number of items be $m$, the basket data can be stored in an information table with $n$ rows and $m$ columns.
Each datum in the data table is boolean to specify whether or not an item is included in the respective transaction.
This is why the rules are called boolean association rules.

The Apriori \cite{AgrawalR1993Mining,AgrawalR1994Apriori} algorithm is based on the Apriori property \cite{AgrawalR1994Apriori}.
It can mine all boolean association rules efficiently given the threshold of support and confidence.
The FP-growth \cite{HanPeiYin00FP} algorithm avoids candidate generation and therefore save computation time further.

  %
  %
  \subsubsection{Quantitative association rules}\label{subsubsection: quantitative-association-rules}
Quantitative association rule \cite{SrikantR1996Mining} was introduced to cope with data tables with quantitative attribute values.
From the data type point of view, it is a generalization of the boolean association rule.
It reveals the relationships among attribute values of an object.
A well known application is mining information of people.
An example of such rule is ``10\% of married people between age 50 and 60 have at least 2 cars; 3\% of all people queried satisfy this rule" \cite{SrikantR1996Mining}.
Similar to the case of boolean association rules, here 10\% is called the \emph{confidence} of the rule, and 3\% the \emph{support} of the rule.

Since the Apriori property still holds in the new context, the Apriori algorithm can be designed accordingly \cite{SrikantR1996Mining}.
One can also follow the idea of FP-growth to design a more efficient algorithm.

  %
  %
  \subsubsection{Multi-level association rules}\label{subsubsection: multi-level-association-rules}
Multi-level association rules \cite{HanFu1995Discovery} reside at multiple concept levels to discover more specific and concrete knowledge from data.
In addition to the transaction data, it requires a description table to indicate different levels.
Suppose that category, content and brand represent the first, the second, and the third level concept respectively of a food.
Two examples of such rules are ``75\% of people buy wheat bread if they buy 2\% milk," and ``82\% of people buy bread if they buy 2\% milk."
However, the rule ``60\% of people buy products made of wheat if they buy 2\% milk" is invalid since
``products made of wheat" does not indicate the category.

  %
  %
  \subsection{Relational association rules}\label{section: relational-association-rules}
In recent years, multi-relational data mining (MRDM) \cite{DzeroskiS2003Multi}, also called relational data mining (RDM), has been proposed to look for patterns that involve multiple tables.
Accordingly, the concept of association rule has been extended with this regard to form relational association rules.
There are various extensions, and we will discuss more popular ones.

  %
  %
  \subsubsection{Extended boolean association rules}\label{subsubsection: extended-boolean-rule}
Dehaspe et al. \cite{DehapseL1998Finding,DehaspeD1999Discovery}, D\v{z}eroski et al. \cite{DzeroskiS2003Multi,DzeroskiS2001Relational}, and Afrati et al. \cite{AftratiF2012Chains} considered the case where binary relations can be chained to produce ternary relations, quaternary relations, etc.
Suppose there are two binary relations, namely the parent-child relation and child-pet relation.
A parent-child-pet relation can be produced using a SQL query on the database.
An example of such rule is ``if a person has a child, then this child has a pet with a probability of 30\%; 20\% of all people satisfy this rule."
Here 30\% is called the \emph{confidence} of the rule, and 20\% is the \emph{support} of the rule.

We will call this type of rules \emph{extended boolean association rules} since they can be viewed a direct extension of boolean association rules on a single table.
The information carried by such rules is limited.
They cannot indicate the number of children a person has, or the number of pets a child has.
Nor can they specify other information, such as the age, of a parent or a child.

Dehaspe et al. \cite{DehaspeD1999Discovery} designed a general purpose inductive logic programming algorithm called {\sc Warmr} to mine this type of rules.
Afrati et al. \cite{AftratiF2012Chains} also tried to attack this problem using integer programming and graph approaches.

  %
  %
  \subsubsection{Decentralized association rules}\label{subsubsection: decentralized-rules}
Jensen et al. \cite{Jensen2000Frequent} considered the case of decentralize tables.
In this case the database contains $n$ primary tables (i.e., tables with one primary key), and one central relationship table (i.e., a table with $n$ foreign keys).
An example of such rule is ``if the ATM type is drive, then the age of the customer is between 20 and 29."
The computation of the confidence and support measures is the same as the table joined from all $n + 1$ tables.

We will call this type of rules \emph{decentralized association rules}.
In fact, if $n = 2$, the database represents a many-to-many relation, which is typical.
However, in real applications a central relationship table seldom exists for $n > 2$.
Therefore these rules are valid for very special databases, or parts of a database.

  %
  %
  \subsubsection{Simple conjunctive association rules}\label{subsubsection: conjunctive-rules}
Goethals et al. \cite{Goethals2008Mining} considered mining association rules in arbitrary relational databases.
This approach looks for pairs of SQL queries $Q_1$ and $Q_2$, such that ``$Q_1$ asks for a set of tuples satisfying a certain condition and $Q_2$ asks for those tuples satisfying a more specific condition" \cite{Goethals2008Mining}.
When the number of tuples matching $Q_2$ is close to that of $Q_1$, a rule is created.
An example of such rule is ``actors starring in `drama' movies typically (with a probability of 90\%) also star in a `comedy' movie."

We will call this type of rules \emph{simple conjunctive association rules}.
The conjunction here is much more flexible than the case of extended boolean association rules.
In fact, any kind of SQL query is supported.
Goethals et al. \cite{Goethals2008Mining} designed the Conqueror algorithm to mine this type of rules.

  %
  %
  \subsubsection{Separated counting association rules}\label{subsubsection: separated-counting-rules}
Goethals et al. \cite{Goethals2010Mining} also considered a more specific type of association rules.
The frequency of a rule is not counted as the number of occurrences in the join of tables.
Let the database consist tables \texttt{Professor}, \texttt{Course} and \texttt{Student}.
For one particular kind of courses, the number of professors who teach them and the number of students who study them are counted separately.
An example of such rule is ``75\% professors named Jan teach courses with 10 credits, among 30\% of all courses."
Here 75\% is the \emph{confidence} and 30\% is the \emph{relative support}.

We will call this type of rules \emph{separated counting association rules}.
Unfortunately, the counting mechanism is limited.
For example, a professor may teach only one course with 10 credits, or she may teach all courses with 10 credits.
This type of rules does not contain such information.
As the association rule becomes more complex in the context of RDM, the support and confidence measures are insufficient to evaluate the strength of the rule.

  %
  %
  \section{Granular association rules with three subtypes}\label{section: granular-rules}
In this section, we will introduce granular association rules to address the drawbacks of existing types mentioned in the last section.
We will first discuss the data model for the new type.
Then we present three subtypes of rules and one general case corresponding to four different explanations of granular association rules.
At the same time, a number of measures are proposed to evaluate the quality of these rules.
A comprehensive comparison with the existing types will be made at the end of the section.

  %
  %
  \subsection{The data model}\label{subsection: data-model}
First we revisit the definitions of information system and binary relation.
At the same time we discuss granules in information systems.
\begin{defn}\label{defn: ins}
$S = (U, A)$ is an information system, where $U = \{x_1, x_2, \dots, x_n\}$ is the set of all objects, $A = \{a_1, a_2, \dots, a_m\}$ is the set of all attributes, and $a_j(x_i)$ is the value of $x_i$ on attribute $a_j$ for $i \in [1..n]$ and $j \in [1..m]$.
\end{defn}

An example of information system is given by Table \ref{subtable: customer}, where $U$ = \{c1, c2, c3, c4, c5\}, and $A$ = \{Age, Gender, Married, Country, Income, NumCars\}.
Another example is given by Table \ref{subtable: product}.
\begin{table}[tb]\caption{A many-to-many entity-relationship system}\label{table: mmer}
\centering
\setlength{\tabcolsep}{5pt}
\subtable[Customer]{
\begin{tabular}{cccccccc}
\hline
CID     & Name     &  Age    & Gender &  Married &  Country     &  Income    & NumCars\\
\hline
c1      & Ron      &  20..29 & Male   &  No      &  USA         &  60k..69k  & 0..1\\
c2      & Michelle &  20..29 & Female &  Yes     &  USA         &  80k..89k  & 0..1\\
c3      & Shun     &  20..29 & Male   &  No      &  China       &  40k..49k  & 0..1\\
c4      & Yamago   &  30..39 & Female &  Yes     &  Japan       &  80k..89k  & 2\\
c5      & Wang     &  30..39 & Male   &  Yes     &  China       &  90k..99k  & 2\\
\hline
\end{tabular}
\label{subtable: customer}
}
\qquad
\setlength{\tabcolsep}{13.4pt}
\subtable[Product]{
\begin{tabular}{ccccccc}
\hline
PID & Name    &  Country    &  Category &  Color   &  Price\\
\hline
p1  & Bread   &  Australia  &  Staple   &  Black   &  1..9   \\
p2  & Diaper  &  China      &  Daily    &  White   &  1..9   \\
p3  & Pork    &  China      &  Meat     &  Red     &  1..9   \\
p4  & Beef    &  Australia  &  Meat     &  Red     &  10..19 \\
p5  & Beer    &  France     &  Alcohol  &  Black   &  10..19 \\
p6  & Wine    &  France     &  Alcohol  &  White   &  10..19 \\
\hline
\end{tabular}\label{subtable: product}
}
\qquad
\setlength{\tabcolsep}{16.4pt}
\subtable[Buys]{
\begin{tabular}{cccccccc}
\hline
CID$\diagdown$ PID &  p1     &  p2     &  p3    & p4      & p5      & p6\\
\hline
c1      &  1      &  1      &  0     &  1      & 1       & 0\\
c2      &  1      &  0      &  0     &  1      & 0       & 1\\
c3      &  0      &  1      &  0     &  0      & 1       & 1\\
c4      &  0      &  1      &  0     &  1      & 1       & 0\\
c5      &  1      &  0      &  0     &  1      & 1       & 1\\
\hline
\end{tabular}
\label{subtable: buys}
}
\end{table}

In an information system, any $A' \subseteq A$ induces an equivalence relation \cite{Pawlak82Rough,SkowronStepaniuk94Approximation}
\begin{equation}\label{equation: equivalence-relation}
E_{A'} = \{(x, y) \in U \times U| \forall a \in A', a(x) = a(y)\},
\end{equation}
and partitions $U$ into a number of disjoint subsets called \emph{blocks} or \emph{granules}.
The block containing $x \in U$ is
\begin{equation}\label{equation: block-contain-x}
E_{A'}(x) = \{y \in U| \forall a \in A', a(y) = a(x)\}.
\end{equation}

The following definition was employed by Yao and Deng \cite{YaoDeng2013Paradigm}.
\begin{defn}\label{defn: granule}
A granule is a triple
\begin{equation}\label{equation: granule}
G = (g, i(g), e(g)),
\end{equation}
where $g$ is the name assigned to the granule, $i(g)$ is a representation of the granule,
and $e(g)$ is a set of objects that are instances of the granule.
\end{defn}

According to Equation (\ref{equation: block-contain-x}), $(A', x)$ determines a granule in an information system.
Hence $g = g(A', x)$ is a natural name to the granule.
$i(g)$ can be formalized as the conjunction of respective attribute-value pairs, i.e.,
\begin{equation}\label{equation: intension-granule}
i(g(A', x)) =  \bigwedge_{a \in A'}\langle a: a(x) \rangle.
\end{equation}
$e(g)$ is given by
\begin{equation}\label{equation: extension-granule}
e(g(A', x)) = E_{A'}(x).
\end{equation}

The \emph{support} of the granule is the size of $e(g)$ divided by the size of the universe, namely,
\begin{equation}\label{equation: support-granule}
supp(g(A', x)) = supp(\bigwedge_{a \in A'}\langle a: a(x) \rangle) = supp(E_{A'}(x)) = \frac{|E_{A'}(x)|}{|U|}.
\end{equation}

In an information system, a granule coincides with a \emph{concept}, which is a basic unit of human thought understood as a pair of intension and extension \cite{SmithMedin1981Categories,YaoDeng2013Paradigm}.
Naturally, $g$, $i(g)$, and $e(g)$ correspond to the name, the intension, and the extension of a concept.
We employ the term \emph{granule} instead of \emph{concept} \cite{MinHuZhu12GranularFour} throughout the paper.

From Equations (\ref{equation: block-contain-x}) and (\ref{equation: extension-granule}), we have the following proposition.
\begin{prop}\label{prop: granule-zoom-in-out}
Let $x \in U$ and $A'' \subset A' \subseteq A$,
\begin{equation}\label{equation: finer-granule}
e(g(A', x)) \subseteq e(g(A'', x)).
\end{equation}
\end{prop}
From the viewpoint of granular computing, we say that $g(A', x)$ is \emph{finer} than $g(A'', x)$, and $g(A'', x)$ is \emph{coarser} than $g(A', x)$.
Proposition \ref{prop: granule-zoom-in-out} show that one can obtain different granules through adding or removing attributes.
Respective operations are called zoom-in and zoom-out, respectively in granular computing.
The technique of obtaining reasonable granules is called feature selection \cite{GiladR2004Margin,JainA1997Feature} in the data mining society, or attribute reduction \cite{MinHeQianZhu11Test,Pawlak82Rough,SkowronRauszer92TheDiscernibility,ZhuWang03Reduction} in the Rough sets society.

\begin{defn}\label{defn: binary-relation}
Let $U = \{x_1, x_2, \dots, x_n\}$ and $V = \{y_1, y_2, \dots, y_k\}$ be two sets of objects.
Any $R \subseteq U \times V$ is a binary relation from $U$ to $V$.
The neighborhood of $x \in U$ is
\begin{equation}\label{equation: relation}
R(x) = \{y \in V | (x, y) \in R\}.
\end{equation}
\end{defn}

When $U = V$ and $R$ is an equivalence relation, $R(x)$ is the equivalence class containing $x$.
From this definition we know immediately that for $y \in V$,
\begin{equation}\label{equation: relation-reverse}
R^{-1}(y) = \{x \in U | (x, y) \in R\}.
\end{equation}

A binary relation is more often stored in the database as a table with two foreign keys.
In this way the storage is saved.
For the convenience of illustration, here we represented it with an $n \times k$ boolean matrix.
An example is given by Table \ref{subtable: buys}, where $U$ is the set of customers as indicated by Table \ref{subtable: customer}, and $V$ is the set of products as indicated by Table \ref{subtable: product}.

With Definitions \ref{defn: ins} and \ref{defn: binary-relation}, we propose the following definition.
\begin{defn}\label{defn: m-m-er}
A many-to-many entity-relationship system (MMER) is a 5-tuple $ES = (U, A, V, B, R)$, where $(U, A)$ and $(V, B)$ are two information systems, and $R \subseteq U \times V$ is a binary relation from $U$ to $V$.
\end{defn}

An example of MMER is given by Table \ref{table: mmer}.

  %
  %
  \subsection{Granular association rules}\label{subsection: different-types}
A \emph{granular association rule} is an implication of the form
\begin{equation}\label{equation: granular-association}
(GR): \bigwedge_{a \in A'}\langle a: a(x) \rangle \Rightarrow \bigwedge_{b \in B'}\langle b: b(y) \rangle,
\end{equation}
where $A' \subseteq A$ and $B' \subseteq B$.

According to Equation (\ref{equation: support-granule}), the set of objects meeting the left-hand side of the granular association rule is
\begin{equation}\label{equation: left-granular-rule}
LH(GR) = E_{A'}(x);
\end{equation}
while the set of objects meeting the right-hand side of the granular association rule is
\begin{equation}\label{equation: right-granular-rule}
RH(GR) = E_{B'}(y).
\end{equation}
We define two measures to evaluate the generality of the granular association rule.
The \emph{source coverage} of $GR$ is
\begin{equation}\label{equation: source-coverage}
scov(GR) = \frac{|LH(GR)|}{|U|};
\end{equation}
while the \emph{target coverage} of $GR$ is
\begin{equation}\label{equation: target-coverage}
tcov(GR) = \frac{|RH(GR)|}{|V|}.
\end{equation}
In most cases, rules with higher source coverage and target coverage tend to be more interesting.
We present a granular association rule for discussion.
\begin{equation}\label{equation: granular-association-rule-general}
\begin{array}{cc}
\langle\textrm{Gender: Male}\rangle \Rightarrow \langle\textrm{Category: Alcohol}\rangle\\
{[}scov = 60\%, tcov = 33\%{]}.
\end{array}
\end{equation}

\begin{figure}[tb]
    \begin{center}
    \includegraphics[width=5in]{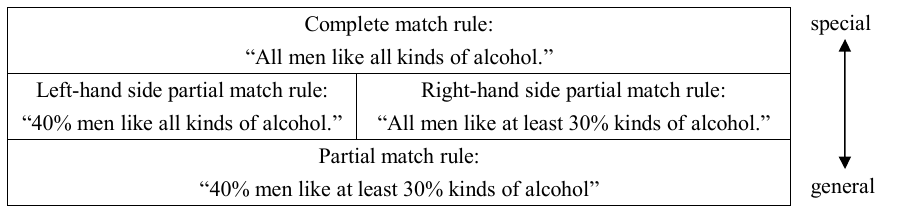}
    \caption{Four explanations of ``men like alcohol"}
    \label{figure: menalcohol}
    \end{center}
\end{figure}

A direct explanation of Rule (\ref{equation: granular-association-rule-general}) is ``men like alcohol."
However, this explanation is ambiguous and the following questions may arise:
Do all men like alcohol?
Do men like all kinds of alcohol?
To avoid such ambiguity, more measures of the rule are needed.
We propose four different explanations of this rule, as illustrated in Figure \ref{figure: menalcohol}, and will discuss them from simple ones to more general ones.
Note that exemplary rules discussed in the following context may not comply to the MMER given by Table \ref{table: mmer}.

  %
  %
  \subsubsection{Complete match}\label{subsubsection: complete-match}
The first explanation of Rule (\ref{equation: granular-association-rule-general}) is ``all men like all alcohol," or equivalently, ``100\% men like 100\% alcohol."
This can be formally expressed by the following definition.
\begin{defn}\label{defn: complete-match}
A granular association rule $GR$ is called a \emph{complete match granular association rule} iff
\begin{equation}\label{equation: all-all}
LH(GR) \times RH(GR) \subseteq R.
\end{equation}
\end{defn}

It is also called a \emph{complete match rule} for brevity.
We need to know the percentage of objects in $U$ matching the rule.
It is called the \emph{support} of the rule and defined by
\begin{equation}\label{equation: support-complete-granular}
supp_c(GR) = scov(GR) = \frac{|LH(GR)|}{|U|},
\end{equation}
where the suffix $c$ stands for \emph{complete}.
Although the support is equal to the source coverage, we still define this measure since in other subtypes they are different. Under this context, the rule
\begin{equation}\label{equation: granular-association-rule-complete}
\begin{array}{cc}
\langle\textrm{Gender: Male}\rangle \Rightarrow \langle\textrm{Category: Alcohol}\rangle\\
{[}scov = 60\%, tcov = 33\%{]},
\end{array}
\end{equation}
will be read as ``all men like all kinds of alcohol; 60\% of all people are men; 33\% of all products are alcohol."
Note that Rules (\ref{equation: granular-association-rule-general}) and (\ref{equation: granular-association-rule-complete}) have the same form.
However the explanation of Rule (\ref{equation: granular-association-rule-complete}) causes no ambiguity under the context of complete match.

  %
  %
  \subsubsection{Left-hand side partial match}\label{subsubsection: left-partial-match}
The second explanation of Rule (\ref{equation: granular-association-rule-general}) is ``some men like all alcohol," or equivalently, ``at least one man like 100\% alcohol."
Because ``some" appears on the left-hand side, the rule is called ``left-hand side partial match."
Consequently, we define a subtype of granular association rule as follows.
\begin{defn}\label{defn: left-partial-match-rule}
A granular association rule $GR$ is called a \emph{left-hand side partial match rule} iff
there exists $x \in LH(GR)$ such that
\begin{equation}\label{equation: left-all}
R(x) \supseteq RH(GR).
\end{equation}
\end{defn}

In applications, however, if very few men like all kinds of alcohol, this rule is not quite useful.
We need to know the percentage of men that like alcohol.
The \emph{support} of the rule is
\begin{equation}\label{equation: support-left-partial-granular}
supp_{lp}(GR) = \frac{|\{x \in LH(GR) | R(x) \supseteq RH(GR)\}|}{|U|}.
\end{equation}
In other words, only men that like all kinds of alcohol are counted.
Moreover, the \emph{source confidence} of the rule is
\begin{equation}\label{equation: confidence-left}
sconf_{lp}(GR) = \frac{|\{x \in LH(GR) | R(x) \supseteq RH(GR)\}|}{|LH(GR)|}.
\end{equation}

One may obtain the following rule
\begin{equation}\label{equation: granular-association-rule-left}
\begin{array}{cc}
\langle\textrm{Gender: Male}\rangle \Rightarrow \langle\textrm{Category: Alcohol}\rangle\\
{[}scov = 60\%, tcov = 33\%, sconf_{lp} = 67\%{]},
\end{array}
\end{equation}
which is read as ``67\% men like all kinds of alcohol; 60\% of customers are men; 33\% of products are alcohol."
We deliberately avoid the support measure in this explanation; the reason will be discussed in the next subsection.

  %
  %
  \subsubsection{Right-hand side partial match}\label{subsubsection: right-partial-match}
The third explanation of Rule (\ref{equation: granular-association-rule-general}) is ``all men like some kinds of alcohol," or equivalently, ``100 \% men like at least one kind of alcohol."
Because ``some" appears on the right-hand side, the rule is called ``right-hand side partial match."
Consequently, we define a subtype of granular association rule as follows.
\begin{defn}\label{defn: right-partial-match-rule}
A granular association rule $GR$ is called a \emph{right-hand side partial match rule} iff
$\forall x \in LH(GR)$,
\begin{equation}\label{equation: right-all}
R(x) \cap RH(GR) \neq \emptyset.
\end{equation}
\end{defn}
Similar to the case of complete match, the \emph{support} of the rule is equal to the source coverage.
It is given by
\begin{equation}\label{equation: support-right-partial-granular}
supp_{rp}(GR) = scov(GR) = \frac{|LH(GR)|}{|U|}.
\end{equation}

In the case of complete match and left-hand side partial match, bigger target coverage values indicate stronger rules.
Unfortunately, in the case of right-hand side partial match, bigger target coverage values indicate weaker rules.
Consider one extreme case as follows: ``all customers like at least one kind of all products."
The rule always holds, and both the source coverage and the target coverage of the rule are 100\%, but the rule is totally useless.

Therefore we need to know how many kinds of alcohol men like.
Here we introduce a new measure called \emph{target confidence} for this purpose.
The \emph{target confidence} of the right-hand side partial match rule is
\begin{equation}\label{equation: target-confidence-right}
tconf_{rp}(GR)
= \min_{x \in LH(GR)}\frac{|R(x) \cap RH(GR)|}{|RH(GR)|}.
\end{equation}

With the existing measures, we may obtain the following rule
\begin{equation}\label{equation: granular-association-rule-right}
\begin{array}{cc}
\langle\textrm{Gender: Male}\rangle \Rightarrow \langle\textrm{Category: Alcohol}\rangle\\
{[}scov_{rp} = 60\%, tcov = 33\%, tconf_{rp} = 50\%{]},
\end{array}
\end{equation}
which is read as ``all men like at least 50\% of alcohol; 60\% of customers are men; 33\% of products are alcohol."

  %
  %
  \subsubsection{Partial match}\label{subsubsection: partial-match}
The fourth explanation of Rule (\ref{equation: granular-association-rule-general}) is ``some men like some kinds of alcohol," or equivalently, ``at least one man like at least one kind of alcohol."
Because ``some" appears on both sides, the rule will be simply called ``partial match."
Consequently, we define this type of granular association rule as follows.
\begin{defn}\label{defn: partial-match-rule}
A granular association rule $GR$ is called a \emph{partial match granular association rule} iff
there exists $x \in LH(GR)$ and $y \in RH(GR)$ such that
\begin{equation}\label{equation: partial-match}
(x, y) \in R.
\end{equation}
\end{defn}
It is also called a \emph{partial match rule} for brevity.
According to the definition, partial match is a general case of granular association rules.
Therefore we cannot call it a \emph{subtype}.

There is a tradeoff between the source confidence and the target confidence of a rule.
Consequently, neither value can be obtained directly from the rule.
To compute any one of them, we need to specify the threshold of the other.
Let $tc$ be the target confidence threshold.
The \emph{support} of the partial match rule is
\begin{equation}\label{equation: support-partial-granular}
supp(GR, tc) = \frac{|\{x \in LH(GR) | \frac{|R(x) \cap RH(GR)|}{|RH(GR)|} \geq tc\}|}{|U|}.
\end{equation}
Here $tc$ is a necessary parameter.
Unlike $supp_{lp}(GR, tc)$ and $supp_{rp}(GR, tc)$, we do not use $supp_p(GR, tc)$ since this is the most general case.

For convenience, in some cases we may ignore it to keep the same form as others.
The \emph{source confidence} of the partial match rule is
\begin{equation}\label{equation: confidence-partial-granular}
sconf(GR, tc)
= \frac{|\{x \in LH(GR) | \frac{|R(x) \cap RH(GR)|}{|RH(GR)|} \geq tc\}|}{|LH(GR)|}.
\end{equation}

Let $sc$ be the source confidence threshold, and
\begin{equation}\label{equation: K-boundary}
\begin{array}{ll}
|\{x \in LH(GR) | |R(x) \cap RH(GR)| \geq K + 1\}|\\
< sc \times |LH(GR)|\\
\leq |\{x \in LH(GR) | |R(x) \cap RH(GR)| \geq K\}|.
\end{array}
\end{equation}
This equation means that $sc \times 100\%$ elements in $LH(GR)$ have connections with at least $K$ elements in $RH(GR)$, but less than $sc \times 100\%$ elements in $LH(GR)$ have connections with at least $K + 1$ elements in $RH(GR)$.
The \emph{target confidence} of the partial match rule is
\begin{equation}\label{equation: target-confidence-partial}
tconf(GR, sc) = \frac{K}{|RH(GR)|}.
\end{equation}
In fact, the computation of $K$ is non-trivial.
First, for any $x \in LH(GR)$, we need to compute $tc(x) = |R(x) \cap RH(GR)|$ and obtain an array of integers.
Second, we sort the array in a descending order.
Third, let $k = \lfloor sc \times |LH(GR)|\rfloor$, $K$ is the $k$-th element in the array.

With the existing measures, we may obtain the following rule
\begin{equation}\label{equation: granular-association-rule-partial}
\begin{array}{cc}
\langle\textrm{Gender: Male}\rangle \Rightarrow \langle\textrm{Category: Alcohol}\rangle\\
{[}scov = 60\%, tcov = 33\%, sconf = 40\%, tconf = 30\%{]},
\end{array}
\end{equation}
which is read as ``40\% men like at least 30\% of alcohol; 60\% of customers are men; 33\% of products are alcohol."
Note again that to represent general cases, the example rules may not comply to Table \ref{table: mmer}.
From Table \ref{table: mmer} we obtain ``100\% men like at least 50\% of alcohol" instead.

  %
  %
  \subsection{Discussion of measures}\label{subsection: discussion-measures}
\setlength{\tabcolsep}{5pt}
\begin{table}[tb]\caption{Summary of source confidence and target confidence}
\label{table: summary-measures}
\begin{center}
\begin{tabular}{lcccccc}
\hline
Subtype $\diagdown$ Measure   &  Source confidence   &  Target confidence\\
\hline
Complete match                &  100\% &  100\%  \\
Left-hand side partial match  &  $\frac{|\{x \in LH(GR) | R(x) \supseteq RH(GR)\}|}{|LH(GR)|}$   &  100\%  \\
Right-hand side partial match &  100\% &  $\min_{x \in LH(GR)}\frac{|R(x) \cap RH(GR)|}{|RH(GR)|}$    \\
Partial match                 &  $\frac{|\{x \in LH(GR) | \frac{|R(x) \cap RH(GR)|}{|RH(GR)|} \geq tc\}|}{|LH(GR)|}$   &  $\frac{K}{|RH(GR)|}$    \\
\hline
\end{tabular}
\end{center}
\end{table}

We have presented five measures to evaluate the quality of granular association rules.
The source coverage is always $\frac{|LH(GR)|}{|U|}$, and the target coverage is always $\frac{|RH(GR)|}{|U|}$.
Table \ref{table: summary-measures} summaries source confidence and target confidence.
From Equations (\ref{equation: support-complete-granular}), (\ref{equation: support-left-partial-granular}), (\ref{equation: confidence-left}), (\ref{equation: support-right-partial-granular}), (\ref{equation: support-partial-granular}) and (\ref{equation: confidence-partial-granular}) we know that for all four cases, there is a direct connection among the support, source coverage and confidence of a rule.
\begin{equation}\label{equation: coverage-support-confidence}
supp_{*}(GR) = scov(GR) \times sconf_{*}(GR),
\end{equation}
where the suffix ``*" could be replaced by $c$, $lp$, and $rp$, or even be removed for the case of partial match.
Hence any one of these three measures can be viewed redundant.
For convenience, in the following context we will ignore the support measure.

  %
  %
  \subsection{Alternative definitions}\label{subsection: alternative}
It is worth noting that Definitions \ref{defn: left-partial-match-rule} and \ref{defn: right-partial-match-rule} are asymmetric.
A symmetric definition of Definition \ref{defn: left-partial-match-rule} is
\begin{defn}\label{defn: alternative-right-partial-match-rule}
A granular association rule $GR$ is called a \emph{type-2 right-hand side partial match  rule} iff
there exists $y \in RH(GR)$ such that
\begin{equation}\label{equation: alternative-left-all}
R^{-1}(y) \supseteq LH(GR).
\end{equation}
\end{defn}

With Definition \ref{defn: alternative-right-partial-match-rule}, we have the following explanation of the rule ``at least one kind of alcohol favors all men."
Moreover, a symmetric definition of Definition \ref{defn: right-partial-match-rule} is
\begin{defn}\label{defn: alternative-left-partial-match-rule}
A granular association rule $GR$ is called a \emph{type-2 left-hand side partial match rule} iff
$\forall y \in RH(GR)$
\begin{equation}\label{equation: alternative-right-all}
R^{-1}(y) \cap LH(GR) \neq \emptyset.
\end{equation}
\end{defn}

With Definition \ref{defn: alternative-left-partial-match-rule}, we have the following explanation of the rule ``all kinds of alcohol favors at least one man."
Unfortunately, the subject these new rules are granules in $V$, and the relation under consideration is $R^{-1}$ instead of $R$.
Therefore the alternative definitions are not appropriate for our situation.

  %
  %
  \subsection{Comparison with the existing types}\label{subsection: comparison-types}
In this subsection, we first analyze the position of granular association rules.
Then we compare the new type with the existing types in more detail.
Finally we discuss its relationships with recommender systems.
These discussions may clarify the motivation of the new type of rules.

Let us draw a road map of association rule development as follows.
\textbf{Boolean association rules} $\rightarrow$ \textbf{quantitative association rules} $\rightarrow$ \textbf{association rules on two universes}
$\rightarrow$ \textbf{association rules on relational databases}.
As discussed in Section \ref{subsubsection: boolean-association-rules}, boolean association rules are the simplest type involving only one universe and only boolean values.
As discussed in Section \ref{subsubsection: quantitative-association-rules}, quantitative association rules also involves one universe, however the data are quantitative.
It is a natural generalization of the boolean association rule.
Granular association rules proposed in the paper involve exactly two universes.
Multi-relational association rules, as discussed in Section \ref{section: relational-association-rules}, may involve databases with tens of universe/relations.
In other words, granular association rules on two universes fill the gap between quantitative association rules and general relational association rules.
We argue that this particular type of rule is important since the many-to-many entity relationship is fundamental in databases.

We now compare granular association rules with other types of association rules mentioned in Section \ref{section: related-works}.
\begin{enumerate}
\item{Both boolean association rules and granular association rules deal with binary relations on two universes.
For granular association rules, objects in either universe are described by a number of attributes.
Therefore granular association rules reveal connections between object subsets (granules) in two universes, while boolean association rules reveal connections between objects in one universe.}

\item{Both quantitative association rules and granular association rules deal with quantitative data.
Moreover, the data sources are all described by attributes.
Quantitative association rules involve only one universe, while granular association rules always involve two.}

\item{Both multi-layer association rules and granular association rules describe objects with attributes.
Multi-layer association rules have a predefined concept/granule hierarchy with a tree structure, which does not exist for granular association rules.
Moreover, Multi-layer association rules involve only one universe.}

\item{Extended boolean association rules may involve more than two data tables.
Similar to boolean association rules, objects are not described by attributes.
Therefore they reveal connections between objects in different universes.}

\item{Decentralized association rules involve at least two primary tables.
From this viewpoint, they are more general than granular association rules.
As mentioned earlier, this type of rules have a special requirement on the database.
Hence they are less useful than granular association rules.}

\item{Simple conjunctive association rules are quite flexible.
They reveal the connections between a object set and one of its subsets.
And the motivation is totally different from granular association rules.}

\item{Granular association rules have the same form as separated counting association rules.
The number of objects for a rule is counted locally in one universe; therefore the joining of tables is unnecessary.
One important difference between two types lies in that granular association rules have more measures, therefore they are semantically richer.}
\end{enumerate}

Now we compare granular association rules and separated counting association rules through an example.
From Table \ref{table: mmer} we obtain the following rule:
\begin{equation}\label{equation: separate-counting-rule}
\{(\textrm{Customer.Gender = Male})\} \Rightarrow_\textrm{Product.PID} \{(\textrm{Product.Category = Alcohol})\}
\end{equation}
The \emph{confidence} of the rule is 2/5 = 40\% since there are 2 kinds of products \{p5, p6\} bought by men that are alcohol, compared to the 5 products \{p1, p2, p4, p5, p6\} bought by men in total.
The \emph{relative support} is 2/6 = 33.3\% since there are 6 products in total.
Here we observe Rules  (\ref{equation: granular-association-rule-partial}) and (\ref{equation: separate-counting-rule}) have quite similar forms, however their measures are totally different.

There is still another closely related technique called collaborative recommendation \cite{BalabanovicM1997Fab} or collaborative filtering \cite{Goldberg1992usingcollaborative}.
This technique also considers many-to-many relationships with some interesting applications such as product recommending and web page recommending.
The recommendation is personalized, i.e., they consider historical behaviors of a customer.
For example, if a customer buys France red wine today, the system may recommend Australia white wine to her.
In contrast, granular association rules identify customers through basic information such as gender, age, country.
Therefore they can be employed to deal with the cold-start problem \cite{ScheinAI2002Methods,SuXY2009Collaborative}, where the customer or the product has just entered the system.

  %
  %
  \section{Granular association rule mining algorithms}\label{section: algorithms}
In this section, we first define the granular association rule mining problem.
Then we propose a sandwich algorithm with four rule checking approaches, one for partial match rules and three for subtypes.
Naturally, the one for partial match rules is also valid for three subtypes.
Then two more algorithms are designed for the complete match subtype.
Time complexities of all algorithms are analyzed.

  %
  %
  \subsection{The granular association rule mining problem}\label{subsection: problem}
We now define the problem as follows.
\begin{problem}\label{problem: partial-rule}
The granular association rule mining problem.

\textbf{Input:} An $ES = (U, A, V, B, R)$, a minimal source coverage threshold $ms$, a minimal target coverage threshold $mt$, a minimal source confidence threshold $sc$, and a minimal target confidence threshold $tc$.

\textbf{Output:} All granular association rules satisfying $scov(GR) \geq ms$, $tcov(GR) \geq mt$, $sconf(GR) \geq sc$, and $tconf(GR) \geq tc$.
\end{problem}

  %
  %
  \subsection{A sandwich algorithm}\label{subsection: sandwich}
A straightforward algorithm for Problem \ref{problem: partial-rule} is given by Algorithm \ref{algorithm: sandwich}. It essentially has three steps.

\textbf{Step 1.} Search in $(U, A)$ all granules meeting the minimal source coverage threshold $ms$.
This step corresponds to Line 1 of the algorithm, where $SG$ stands for source granule.

\textbf{Step 2.} Search in $(V, B)$ all granules meeting the minimal target coverage threshold $mt$.
This step corresponds to Line 2 of the algorithm, where $TG$ stands for target granule.

\textbf{Step 3.} Check all possible rule regarding $SG$ and $TG$, and output valid ones.
This step corresponds to Lines 3 through 10 of the algorithm.

Since this algorithm starts from both ends of the association rule and proceeds to the middle, it is called the ``sandwich" algorithm.
Note that the check of the condition $sconf(GR, tc) \geq sc$ in Line 6 is non-trivial.
And it indicates both thresholds of source confidence and target confidence should be met.

\begin{algorithm}[tb!]\caption{A sandwich algorithm for partial match}\label{algorithm: sandwich}
  \textbf{Input}: $ES = (U, A, V, B, R)$, $ms$, $mt$, $sc$, $tc$.\\
  \textbf{Output}: All partial match rules satisfying given constraints.\\
  \textbf{Method}: partial-match-sandwich\\
  \begin{algorithmic}[1]
    \STATE $SG(ms) = \{(A', x) \in 2^A \times U| \frac{|E_{A'}(x)|}{|U|} \geq ms\}$;
    \STATE $TG(mt) = \{(B', y) \in 2^B \times V| \frac{|E_{B'}(y)|}{|V|} \geq mt\}$;
    \FOR {each $g \in SG(ms)$}
      \FOR {each $g' \in TG(mt)$}
        \STATE $GR = (i(g) \Rightarrow i(g'))$;
        \IF {$sconf(GR, tc) \geq sc$}
          \STATE output rule $GR$;
        \ENDIF
      \ENDFOR
    \ENDFOR
  \end{algorithmic}
\end{algorithm}

Now we discuss the algorithm in more detail.
The Apriori algorithm \cite{AgrawalR1994Apriori,SrikantR1996Mining} and the FP-growth algorithm \cite{HanPeiYin00FP} can be employed in Lines 1 and 2.
These algorithms are based on the Apriori property, which is stated as ``every subset of a frequent itemset must also be a frequent itemset" \cite{AgrawalR1994Apriori}.
Under our context, the Apriori property can be restated as follows.
\begin{property}\label{property: apriori}
Let $A'' \subset A' \subseteq A$ and $x \in U$.
\begin{equation}\label{equation: apriori}
|E_{A'}(x)| \leq |E_{A''}(x)|.
\end{equation}
\end{property}

Naturally, for three subtypes, the condition expressed by Line 6 of the algorithm might be replaced by simpler ones.
We will explain the cases for each subtypes.
  %
  %
  \subsubsection{Complete match}\label{subsubsection: complete-match-algorithm}
If $sc = tc  = 100\%$, we are essentially looking for complete match rules.
The condition can be replaced by
\begin{equation}\label{equation: condition-complete-match}
e(g) \times e(g') \subseteq R.
\end{equation}

Moreover, in this case some checks are redundant.
We have the following property.
\begin{property}\label{property: apriori-two-sides}
Let $A'' \subset A' \subseteq A$, $x \in U$, $B'' \subset B' \subseteq B$, and $y \in V$.
If $e(g(A'', x)) \times e(g(B'', y)) \subseteq R$,
\begin{equation}\label{equation: apriori-two-sides}
e(g(A', x)) \times e(g(B', y)) \subseteq R.
\end{equation}
\end{property}
\begin{pf}
Because $A'' \subset A'$, $e(g(A', x)) \subseteq e(g(A'', x))$.
Similarly $e(g(B', y)) \subseteq e(g(B'', y))$.
Therefore $e(g(A', x)) \times e(g(B', y)) \subseteq e(g(A'', x)) \times e(g(B'', y))$.
And the property holds.
\end{pf}

Property \ref{property: apriori-two-sides} is essentially another form of the Apriori property.
Its converse negative proposition can be used to remove unnecessary check of rules.
Note that changes can be made on both sides of the rule.
For example, if rule ``all Chinese men like all kinds of France alcohol" does not hold, then rule ``all men like all kinds of alcohol" never holds.

  %
  %
  \subsubsection{Left-hand side partial match}\label{subsubsection: left-partial-match-algorithm}
If $tc  = 100\%$, we are essentially looking for left-hand side partial match rules.
The condition can be replaced by
\begin{equation}\label{equation: condition-left-partial-match}
\frac{|\{x \in LH(GR) | R(x) \supseteq RH(GR)\}|}{|LH(GR)|} \geq sc.
\end{equation}

  %
  %
  \subsubsection{Right-hand side partial match}\label{subsubsection: right-partial-match-algorithm}
If $sc  = 100\%$, we are essentially looking for right-hand side partial match rules.
The condition can be replaced by
\begin{equation}\label{equation: condition-right-partial-match}
\min_{x \in e(g)}\frac{|R(x) \cap e(g')|}{|e(g')|} \geq tc.
\end{equation}

Similar to the case of complete match, we would like to remove unnecessary check of rules.
In fact, we have the following property.
\begin{property}\label{property: apriori-right-side}
Let $A'' \subset A' \subseteq A$, $x \in U$, $B' \subseteq B$, and $y \in V$.
\begin{equation}\label{equation: apriori-right-side}
\min_{x' \in e(g(A', x))}\frac{|R(x) \cap e(g(B', y))|}{|e(g(B', y))|}
\geq \min_{x' \in e(g(A'', x))}\frac{|R(x) \cap e(g((B', y))|}{|e(g(B', y))|}.
\end{equation}
\end{property}
\begin{pf}
Because $A'' \subseteq A'$, $e(g(A'', x)) \supseteq e(g(A', x))$.
Hence Equation (\ref{equation: apriori-right-side}) holds.
\end{pf}

Property \ref{property: apriori-right-side} indicates one approach to removing unnecessary check concerning the left side of the rule.
Unlike Property \ref{property: apriori-two-sides}, in this case the change cannot be made on both sides.
For example, if rule ``all Chinese men like at least 30\% kinds of France alcohol" does not hold, then ``all men like at least 30\% kinds of France alcohol" never holds.
However, ``all Chinese men like at least 30\% kinds of alcohol" may hold.

Now we analyze the time complexity of the algorithm.
For the partial match subtype, from Equation (\ref{equation: support-partial-granular}) we know the complexity of Line 6 is
\begin{equation}\label{equation: granule-complexity}
O(|e(g)| \times |e(g')|) = O(|U| \times |V|).
\end{equation}
According to the \textbf{for} loops, the time complexity of Algorithm \ref{algorithm: sandwich} is
\begin{equation}\label{time-complexity-sandwich}
O(|SG(ms)| \times |TG(mt)| \times |U| \times |V|).
\end{equation}

For the complete match subtype, suppose that both $e(g)$ and $e(g')$ are stored in 1-dimensional positive number arrays.
Each element in the array indicates the inclusion of one particular object in the granule.
For example, $[1, 4, 8]$ indicates $\{x_1, x_4, x_8\}$.
Suppose further that $R$ is stored in a $|U| \times |V|$ boolean array.
The time complexity of checking $e(g) \times e(g') \subseteq R$ is the same as that of partial match as indicated by Equation (\ref{equation: granule-complexity}).
Consequently, this time complexity for the complete match subtype is also given by Equation (\ref{time-complexity-sandwich}).
However, checking $e(g) \times e(g') \subseteq R$ ends immediately once a violation of the relationship is found.
Compared with the check of $sconf(GR, tc) \geq sc$, it is less time consuming.

Similarly, for the other two subtypes, the time complexities are all given by Equation (\ref{time-complexity-sandwich}).
The run time for different subtypes will, however, be very different in applications.
This will be shown through experiments in Section \ref{section: experiments}.

  %
  %
  \subsection{Two algorithms for the complete match subtype}\label{subsection: approaches-complete-match}
The time complexity of the sandwich algorithm is quite high.
Now we propose two alternative approaches for the complete match subtype.
We will show that their time complexities are lower than Algorithm \ref{algorithm: sandwich}.

  %
  %
  \subsubsection{A forward algorithm}\label{subsubsection: forward}
\begin{algorithm}[tb!]\caption{A forward algorithm}\label{algorithm: complete-rule-mining-forward}
  \textbf{Input}: $ES = (U, A, V, B, R)$, $ms$, $mt$.\\
  \textbf{Output}: All complete match granular association rules satisfying given constraints.\\
  \textbf{Method}: complete-match-rules-forward\\
  \begin{algorithmic}[1]
    \STATE $SG(ms) = \{(A', x) \in 2^A \times U | \frac{|E_{A'}(x)|}{|U|} \geq ms\}$;
    \STATE $TG(mt) = \{(B', y) \in 2^B \times V | \frac{|E_{B'}(y)|}{|V|} \geq mt\}$;
    \FOR {each $g \in SG(ms)$}
      \STATE $X = e(g)$;
      \STATE $Y = \underline{R}(X)$;
      \FOR {each $g' \in TG(mt)$}
        \IF {($e(g') \subseteq Y$)}
          \STATE output rule $i(g) \Rightarrow i(g')$;
        \ENDIF
      \ENDFOR
    \ENDFOR
  \end{algorithmic}
\end{algorithm}

The first alternative approach is called the ``forward" approach.
It starts from the left-hand side of the rule and proceeds to the right-hand side.
The algorithm is listed in Algorithm \ref{algorithm: complete-rule-mining-forward}.
It essentially has four steps.

\textbf{Steps 1 and 2.} They are the same as Algorithm \ref{algorithm: sandwich}.

\textbf{Step 3.} For each granule obtained in Step 1, construct a block in $V$ according to $R$.
This step corresponds to Line 4 of the algorithm.
The function $e$ has been defined in Equation (\ref{equation: extension-granule}).
We introduce a new concept regarding Line 5.
\begin{defn}\label{defn: lower-approximation}
Let $U$ and $V$ be two universes, $R \subseteq U \times V$ be a binary relation, $X \subseteq U$.
The lower approximation of $X$ with respect to $R$ is
\begin{equation}\label{equation: forward-neighbor-lower-approximation}
\underline{R}(X) = \{y \in V | R^{-1}(y) \supseteq X\}.
\end{equation}
\end{defn}
In our example, $\underline{R}(X)$ are all products that favor all people in $X$.
The concept ``lower approximation" comes from rough sets \cite{Pawlak82Rough}.
However, we consider two universes here instead of only one.

\textbf{Step 4.} Check possible rules regarding $C'$ and $Y$, and output all rules.
This step corresponds to Lines 6 through 10 of the algorithm.
In Line 7, since $e(g')$ and $Y$ could be stored in sorted arrays, the complexity of checking $e(g') \subseteq Y$ is
\begin{equation}\label{equation: addition-complexity}
O(|e(g')| + |Y|) = O(|V|).
\end{equation}
According to the \textbf{for} loops, the time complexity of Algorithm \ref{algorithm: complete-rule-mining-forward} is
\begin{equation}\label{time-complexity-forward}
O(|SG(ms)| \times |TG(mt)| \times |V|),
\end{equation}
which is lower than Algorithm \ref{algorithm: sandwich}.

  %
  %
  \subsection{A backward algorithm}\label{subsection: backward}
A backward algorithm, which is a dual of Algorithm \ref{algorithm: complete-rule-mining-forward}, is listed in Algorithm \ref{algorithm: complete-rule-mining-backward}.
It starts from the right-hand side of the rule and proceeds to the left-hand side.
It is symmetric with respect to Algorithm \ref{algorithm: complete-rule-mining-forward}.
According to Definition \ref{defn: lower-approximation}, $\underline{R^{-1}}(Y) = \{x \in U | R(x) \supseteq Y\}$.
In our example, $\underline{R^{-1}}(Y)$ are all people buying all products in $Y$.
Similar to the analysis of Algorithm \ref{algorithm: complete-rule-mining-forward}, the time complexity of Algorithm \ref{algorithm: complete-rule-mining-backward} is
\begin{equation}\label{time-complexity-backward}
O(|SG(ms)| \times |TG(mt)| \times |U|).
\end{equation}

\begin{algorithm}[tb!]\caption{A backward algorithm}\label{algorithm: complete-rule-mining-backward}
  \textbf{Input}: $ES = (U, A, V, B, R)$, $ms$, $mt$.\\
  \textbf{Output}: All complete match granular association rules satisfying given constraints.\\
  \textbf{Method}: complete-match-rules-backward\\
  \begin{algorithmic}[1]
    \STATE $SG(ms) = \{(A', x) \in 2^A \times U | \frac{|E_{A'}(x)|}{|U|} \geq ms\}$;
    \STATE $TG(mt) = \{(B', y) \in 2^B \times V | \frac{|E_{B'}(y)|}{|V|} \geq mt\}$;
    \FOR {each $g' \in TG(ms)$}
      \STATE $Y = e(g')$;
      \STATE $X = \underline{R^{-1}}(Y)$;
      \FOR {each $g \in SG(mt)$}
        \IF {($e(g) \subseteq X$)}
          \STATE output rule $i(g) \Rightarrow i(g')$;
        \ENDIF
      \ENDFOR
    \ENDFOR
  \end{algorithmic}
\end{algorithm}

Now one question arises: which algorithm performs better?
According to Equations (\ref{time-complexity-forward}) and (\ref{time-complexity-backward}), we should choose the forward algorithm if $|U| < |V|$, and the backward algorithm otherwise.
This issue will be discussed further through experimentation in Section \ref{section: experiments}.

  %
  %
  \section{Experiments on real world datasets}\label{section: experiments}
In this section, we try to answer the following problems through experimentation.
\begin{enumerate}
\item{Do granular association rules make sense in real-world applications?}
\item{Do different subtypes of granular association rules exist in real-world applications? If so, how frequent do they occur?}
\item{How does a rule's source confidence and target confidence influence each other?}
\item{Do dedicated approaches for different subtypes improve the performance of the sandwich algorithm?}
\item{Do the forward and backward algorithms outperform the sandwich algorithm significantly?}
\end{enumerate}

  %
  %
  \subsection{Datasets}\label{subsection: datasets}
We tested our algorithms on two real world data sets.
The first data set is MovieLens \cite{movielens} assembled by the GroupLens project \cite{grouplens}.
It is widely used in recommender systems (see, e.g., \cite{Herlocker1999collaborative,ScheinA2002ColdStart}).
The database schema is as follows.
\begin{enumerate}
\item[$\bullet$]{User (\underline{userID}, age, gender, occupation)}
\item[$\bullet$]{Movie (\underline{movieID}, release-year, genre)}
\item[$\bullet$]{Rates (\underline{userID, movieID})}
\end{enumerate}
We use the version with 943 users and 1,682 movies.
The data are preprocessed to cope with Definition \ref{defn: m-m-er} as follows.
The original Rate relation contains the rating of movies with 5 scales, while we only consider whether or not a user has rated a movie.
The user age is discretized to 9 intervals as indicated by the data set.
Since there are few movies before 1970s and too many movies after 1990, the release year is discretized to 3 intervals: before 1970s, 1970s-1980s, and 1990s.
The genre is a multi-valued attribute.
Therefore we scale it to 18 boolean attributes, namely, action, adventure, animation, children, comedy, crime, documentary, drama, fantasy, FilmNoir, horror, musical, mystery, romance, scientific-fiction, thriller, war, and western.

The second data set is general education course selection from Minnan Normal University.
The database schema is as follows.
\begin{enumerate}
\item[$\bullet$]{Student (\underline{studentID}, name, gender, birth-year, politics-status, grade, department, nationality, length-of-schooling)}
\item[$\bullet$]{Course (\underline{courseID}, credit, class-hours, availability, department)}
\item[$\bullet$]{Selects (\underline{studentID, courseID})}
\end{enumerate}
We collected data during the semester between 2011 and 2012.
There are 145 general education courses in the university, and 9,654 students took part in course selection.

  %
  %
  \subsection{Results}\label{subsection: results}
We undertake four sets of experiments to answer the questions raised at the beginning of the section one by one.

  %
  %
  \subsubsection{The meaningfulness of rules}\label{subsubsection: rule-meaning}
First we look at some rules of the MovieLens data set.
The setting is as follows: $ms = 0.08$, $mt = 0.01$, $sc = 0.20$, and $tc = 0.20$.
641 granular association rules are obtained.
4 of them are listed below.

\noindent(Rule 1) $\langle\textrm{gender: M}\rangle \wedge \langle \textrm{occupation: student}\rangle$\\
\indent$\Rightarrow \langle\textrm{year: 1990s}\rangle \wedge \langle \textrm{science-fiction: 1}\rangle \wedge \langle \textrm{thriller: 1}\rangle$\\
\indent$[scov = 0.144, tcov = 0.014, sconf = 0.235, tconf = 0.200]$

\noindent(Rule 2) $\langle\textrm{age: 35..40}\rangle \wedge \langle \textrm{gender: M}\rangle$\\
\indent$\Rightarrow \langle\textrm{year: 1970s - 1980s}\rangle \wedge \langle\textrm{action: 1}\rangle \wedge \langle\textrm{adventure: 1}\rangle$\\
\indent$[scov = 0.132, tcov = 0.010, sconf = 0.448, tconf = 0.200]$

\noindent(Rule 3) $\langle\textrm{age: 25..34}\rangle \wedge \langle \textrm{gender: F}\rangle$\\
\indent$\Rightarrow \langle\textrm{year: 1990s}\rangle \wedge \langle\textrm{action: 1}\rangle \wedge \langle\textrm{romance: 1}\rangle$\\
\indent$[scov = 0.083, tcov = 0.010, sconf = 0.240, tconf = 0.200]$

\noindent(Rule 4) $\langle\textrm{age: 18..24}\rangle \wedge \langle \textrm{gender: M}\rangle \wedge \langle \textrm{occupation: student}\rangle$\\
\indent$\Rightarrow \langle\textrm{animation: 1}\rangle \wedge \langle\textrm{children: 1}\rangle \wedge \langle\textrm{musical: 1}\rangle$\\
\indent$[scov = 0.080, tcov = 0.010, sconf = 0.236, tconf = 0.200]$

Rule 1 indicates that male students would like to watch new movies on both science-fiction and thriller topics.
This rule is stronger than the other three in terms of source coverage and target coverage.
Rule 2 indicates that middle aged men would like to watch movies on both action and adventure topics.
This rule is stronger than the other three in terms of source confidence.
Rule 4 indicates that many young men still like children's cartoons.
All these rules make sense to us.

Second we look at some rules of the course selection data set.
The setting is as follows: $ms = 0.06$, $mt = 0.06$, $sc = 0.18$, and $tc = 0.11$.
40 granular association rules are obtained, and 4 of them listed below.

\noindent(Rule 5) $\langle\textrm{department: economics}\rangle$\\
\indent$\Rightarrow \langle\textrm{department: human-resource}\rangle$\\
\indent$[scov = 0.072, tcov = 0.062, sconf = 0.188, tconf = 0.110]$

\noindent(Rule 6) $\langle\textrm{nationality: han}\rangle \wedge \langle \textrm{department: economics}\rangle$\\
\indent$\Rightarrow \langle\textrm{credit: 1}\rangle \wedge \langle\textrm{department: human-resource}\rangle$\\
\indent$[scov = 0.070, tcov = 0.062, sconf = 0.189, tconf = 0.110]$

\noindent(Rule 7) $\langle\textrm{politics: league-member}\rangle \wedge \langle\textrm{nationality: han}\rangle \wedge \langle\textrm{department: economics} \rangle \wedge$\\ \indent$ \langle\textrm{length-of-schooling: 4} \rangle$\\
\indent$\Rightarrow \langle\textrm{credit: 1}\rangle \wedge \langle\textrm{department: human-resource}\rangle$\\
\indent$[scov = 0.065, tcov = 0.062, sconf = 0.187, tconf = 0.110]$

\noindent(Rule 8) $\langle\textrm{birth-year: 1993}\rangle \wedge \langle\textrm{nationality: han}\rangle \wedge \langle\textrm{length-of-schooling: 4} \rangle \wedge \langle\textrm{grade: 2011}\rangle$\\
\indent$\Rightarrow \langle\textrm{credit: 1}\rangle \wedge \langle\textrm{department: human-resource}\rangle$\\
\indent$[scov = 0.070, tcov = 0.062, sconf = 0.180, tconf = 0.110]$

Rule 5 indicates that students in the economics like courses offered by the human-resource department.
We observe that Rule 7 is finer than Rule 6, which is in turn finer than Rule 5.
It happens that all three rules hold under the given setting.
Rule 8 is not comparable with other three rules in terms of granulation.

Generally, rules mined from the MovieLens data set are more interesting than those mined from the course selection data set.

  %
  %
  \subsubsection{Rules of different subtypes}\label{subsubsection: rule-subtypes}
Different subtypes of granular association rules are one of the major issues of the paper.
We cannot mine any complete match rules from the two data sets.
The reason lies in that complete match rules are too strong.
However, many left(right)-hand side partial match rules can be extracted from both data sets.
Therefore the discussions of subtypes are meaningful.
For brevity only results of the MovieLens data set are presented.

Let $ms = 0.08$, $mt = 0.2$, $sc = 1.0$, $tc = 10^-6$, we obtained 72 right-hand side partial match rules.
One of them is listed below:\\
\noindent(Rule 9) $\langle\textrm{age: 18..24}\rangle \wedge \langle \textrm{gender: M}\rangle \wedge \langle \textrm{occupation: student}\rangle$\\
\indent$\Rightarrow \langle\textrm{year: 1990s}\rangle \wedge \langle \textrm{comedy: 1}\rangle$\\
\indent\indent$[scov = 0.080, tcov = 0.247, sconf = 1.0, tconf = 0.002]$

Let $ms = 0.08$, $mt = 0.01$, $sc = 10^-6$, $tc = 1.0$, we obtained 10 left-hand side partial match rules.
One of them is listed below:\\
\noindent(Rule 10) $\langle\textrm{age: 25..34}\rangle \wedge \langle \textrm{gender: M}\rangle \wedge \langle \textrm{occupation: student}\rangle$\\
\indent$\Rightarrow \langle\textrm{year: 1970s - 1980s}\rangle \wedge \langle \textrm{action: 1}\rangle \wedge \langle \textrm{adventure: 1}\rangle$\\
\indent\indent$[scov = 0.244, tcov = 0.010, sconf = 0.004, tconf = 1.0]$

\begin{figure}[tb]
    \subfigure[]{
    \begin{minipage}[b]{2.55in}
        \centering
        \includegraphics[width=2.5in]{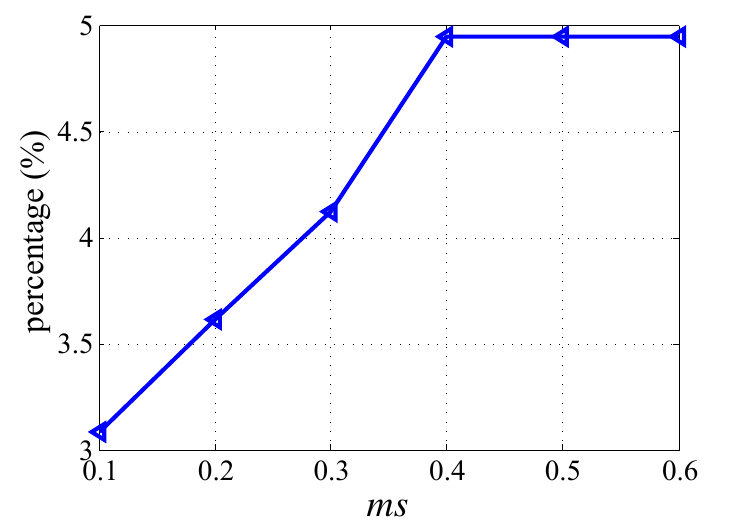}
    \end{minipage}
    }
    \subfigure[]{
    \begin{minipage}[b]{2.55in}
        \centering
        \includegraphics[width=2.5in]{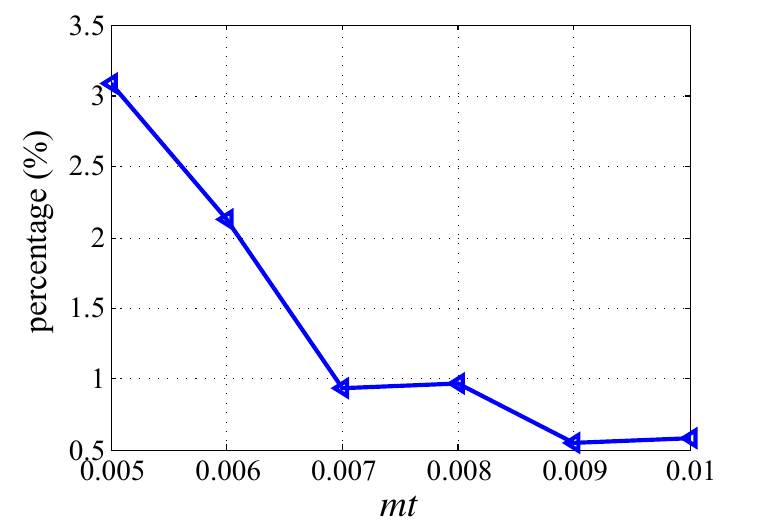}
    \end{minipage}
    }
    \subfigure[]{
    \begin{minipage}[b]{2.55in}
        \centering
        \includegraphics[width=2.5in]{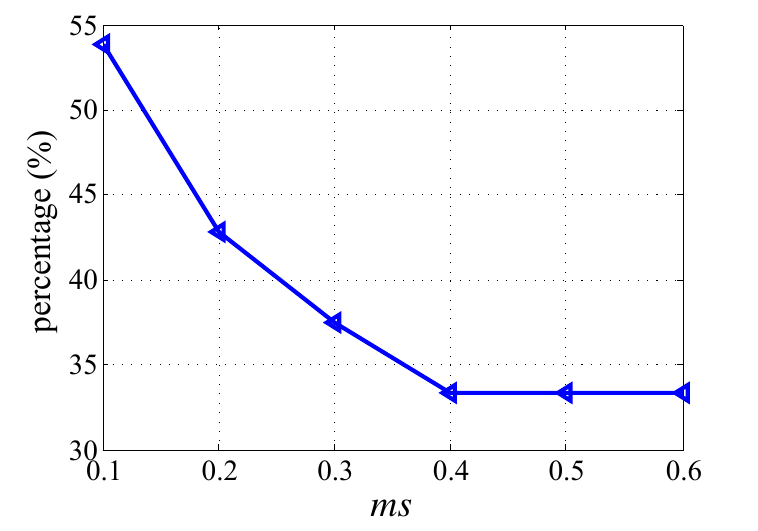}
    \end{minipage}
    }
    \subfigure[]{
    \begin{minipage}[b]{2.55in}
        \centering
        \includegraphics[width=2.5in]{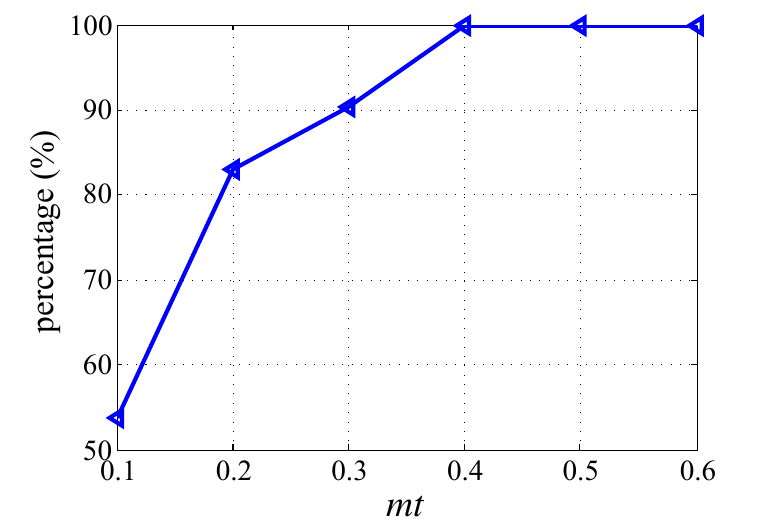}
    \end{minipage}
    }
\caption{The percentage of left(right)-hand side partial match rules: (a)(b) left-hand side partial match, (c)(d) right-hand side partial match}
\label{figure: partial-match-percentage}
\end{figure}

Now we study these rules quantitatively.
Given $ms$ and $mt$, we can compute the set of all source granules $SG(ms)$ and the set of all target granules $TG(mt)$.
The number of all possible rules is $|SG(ms)| \times |TG(mt)|$.
We are interesting in the percentage of rules that can be viewed left(right)-hand side partial match ones.
For this purpose, we set $tc = 1$ ($tc = 10^-6$) and $sc = 10^-6$ ($sc = 1$) to obtain left(right)-hand side partial match rules.
Results are illustrated in Fig \ref{figure: partial-match-percentage}.

Here we observe a number of natural trends.
Figure \ref{figure: partial-match-percentage}(a) shows that given a granule of movies, it is easier to find some users that rate them all in a bigger group.
For example, no male students rate all adventure war movies, however some men do.
Figure \ref{figure: partial-match-percentage}(b) shows that given a granule of users, with the increase of the movie granule, it is harder to find someone who rate all these movies.
For example, some women rate all children musical movies released in 1990s, however none of them rate all movies released in 1990s.
Figure \ref{figure: partial-match-percentage}(c) shows that given a granule of movies, with the increase of the user granule, it is less likely that all these users rate some of these movies.
For example, all men aged between 18 and 25 rate some kind of comedy movies, however not all men do.
Figure \ref{figure: partial-match-percentage}(d) shows that given a granule of users, with the increase of the movie granule, it is more likely that all these users rate some of these movies.
For example, not all male students rate some adventure movies released in 1990s, however they all rate some movies released in 1990s.

Figure \ref{figure: partial-match-percentage} also show that left-hand side partial match rules are infrequent.
This is because there are seldom some people who rate all movies of a kind.
On the contrary, right-hand side partial match rules are frequent because a group of people may rate at least one movie of a kind.
Consider an extreme situation where $mt = 1$.
That is, the target granule is all movies.
Because user rate at least one movie, respective right-hand side partial match rules always hold regardless of the user granule
Figure \ref{figure: partial-match-percentage}(d) shows that the percentage reaches 100\% for $mt \geq 0.4$.

  %
  %
  \subsubsection{The relationship between the source confidence and the target confidence}\label{subsubsection: relationship-confidence}
As discussed in Section \ref{subsubsection: partial-match}, there is a tradeoff between the source confidence and the target confidence of a rule.
Here we discuss this issue in detail through two rules.
The first rule is extracted from the MovieLens data set.\\
\noindent(Rule 11) $\langle\textrm{age: 18..24}\rangle \wedge \langle\textrm{gender: M}\rangle$\\
\indent$\Rightarrow \langle\textrm{action: 1}\rangle  \wedge \langle\textrm{adventure: 1}\rangle \wedge \langle\textrm{science-fiction: 1}\rangle$\\
\indent$[scov = 0.151, tcov = 0.016]$\\
which can be read as ``Male students between 18 to 24 years old like movies that belong to action, adventure and scientific fiction at the same time."
The source coverage and the target coverage of the rule are already indicated.
We set the target confidence threshold to obtain different source confidences.
The target confidence thresholds are set to $10^{-6}$, 0.1, 0.2, \dots, 0.9.
A very small however non-zero (e.g., $10^{-6}$) number guarantees that at least one object is covered by the rule.
We cannot set the threshold as 0 which is meaningless.

\begin{figure}[tb]
    \subfigure[]{
    \begin{minipage}[b]{2.55in}
        \centering
        \includegraphics[width=2.5in]{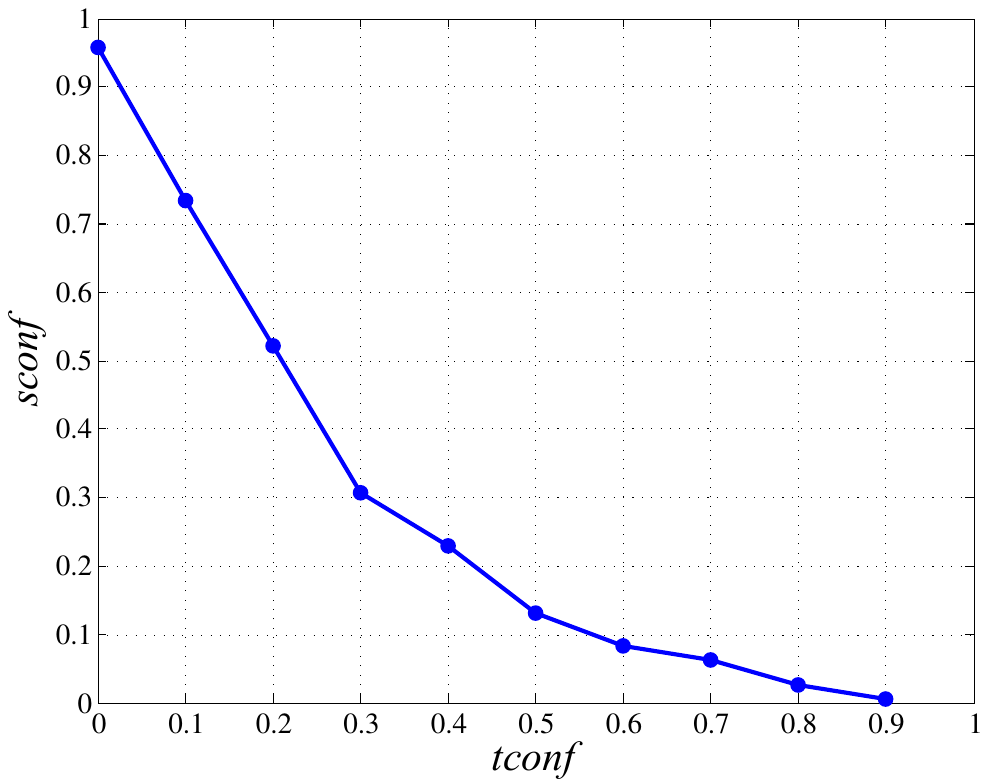}
    \end{minipage}
    }
    \subfigure[]{
    \begin{minipage}[b]{2.55in}
        \centering
        \includegraphics[width=2.5in]{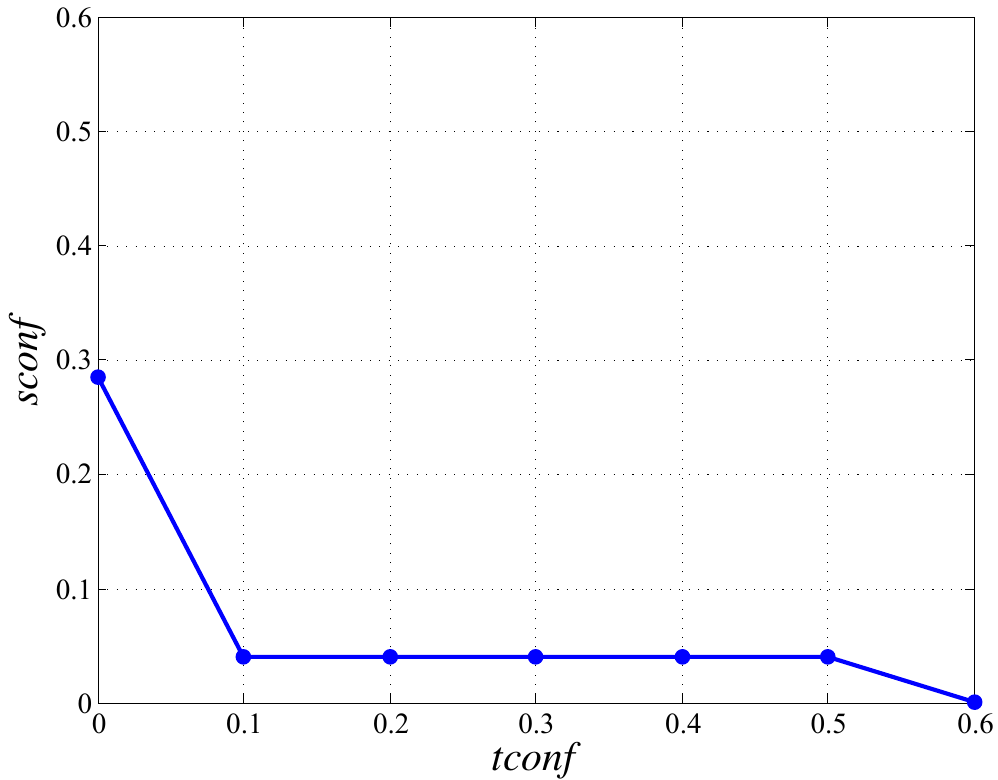}
    \end{minipage}
    }
\caption{The tradeoff between the source and target confidences: (a) Rule 11, (b) Rule 12}
\label{figure: confidence-tradeoff}
\end{figure}

The relationship between the source confidence and the target confidence is depicted in Figure \ref{figure: confidence-tradeoff}(a).
With the increase of the target confidence threshold, the source confidence decreases as we expect.
There are two extreme cases.
One is for $tconf = 10^{-6}$, and the other is for $tconf = 0.9$.
They correspond to two explanation of Rule 11.
The first is ``98\% male students between 18 to 24 years old like at least one movie that belong to action, adventure and scientific fiction at the same time.
While the second is ``at least one male students between 18 to 24 years old like at least 90\% movies that belong to action, adventure and scientific fiction at the same time.
Unfortunately, these extreme cases are not quite useful in recommender systems, and general situations are more interesting.

The second rule is extracted from the course selection data set.\\
\noindent(Rule 12) $\langle\textrm{gender: F}\rangle \wedge \langle\textrm{nationality: han}\rangle \wedge \langle\textrm{length-of-schooling: 4}\rangle$\\
\indent$\Rightarrow \langle\textrm{credit: 1}\rangle \wedge \langle\textrm{department: human-resource}\rangle \wedge \langle\textrm{category: public}\rangle$\\
\indent$[scov = 0.613, tcov = 0.013]$

The relationship between the source confidence and the target confidence is depicted in Figure \ref{figure: confidence-tradeoff}(b).
There is also a tradeoff between these two measures.
Figure \ref{figure: confidence-tradeoff}(a) indicates that the target confidence changes smoothly with the change of the source confidence.
However, Figure \ref{figure: confidence-tradeoff}(b) indicates the change is abrupt.
According to the observation of many other rules, we found that the phenomenon is due to the characteristics of the data set.

  %
  %
  \subsubsection{The performance of dedicated rule checking approaches}\label{subsubsection: performance-dedicated}
We study the performance of the sandwich algorithm for different subtypes.
We focus on Step 3 of the algorithm since it is most time consuming than Steps 1 and 2 for large datasets, and it is different for subtypes.
The algorithm chooses the appropriate subtype according to $sc$ and $tc$ settings, as indicated in Section \ref{subsection: sandwich}.
We deliberately set $sc$ and/or $tc$ to 0.95 such that different subtypes are chosen, while the rule set is the same as the cases of $sc = 1$ and/or $tc = 1$.

\setlength{\tabcolsep}{3pt}
\begin{table}[tb]\caption{Run time of Step 3 for different settings on the course selection data set}
\label{table: step3-results}
\begin{center}
\begin{tabular}{ccccllrr}
\hline
$ms$ & $mt$ & $|SG(ms)|$ & $|TG(mt)|$ & $sc$ & $tc$     & basic operations & run time (ms)\\
\hline
     &      &            &     & 1          & 1         & 30,214           & 0\\
0.06 & 0.06 &  670       & 45  & 0.95       & 1         & 2,479,057        & 31\\
     &      &            &     & 1          & 0.95      & 904,500          & 0\\
     &      &            &     & 0.95       & 0.95      & 1,462,868,100    & 8,594\\
\hline
     &      &            &     & 1          & 1         & 44,999           & 0\\
0.05 & 0.05 &  817       & 55  & 0.95       & 1         & 3,251,891        & 31\\
     &      &            &     & 1          & 0.95      & 1,168,310        & 0\\
     &      &            &     & 0.95       & 0.95      & 1,662,924,120    & 10,375\\
\hline
     &      &            &     & 1          & 1         & 95,371           & 0\\
0.04 & 0.04 &  1,041     & 91  & 0.95       & 1         & 5,860,127        & 31\\
     &      &            &     & 1          & 0.95      & 1,722,855        & 16\\
     &      &            &     & 0.95       & 0.95      & 2,085,064,990    & 13,156\\
\hline
     &      &            &     & 1          & 1         & 259,596          & 0\\
0.03 & 0.03 &  2,268     & 113 & 0.95       & 1         & 9,630,946        & 63\\
     &      &            &     & 1          & 0.95      & 4,003,020        & 31\\
     &      &            &     & 0.95       & 0.95      & 2,925,501,620    & 17,547\\
\hline
     &      &            &     & 1          & 1         & 1,020,287        & 16\\
0.02 & 0.02 &  5,385     & 187 & 0.95       & 1         & 23,211,961       & 156\\
     &      &            &     & 1          & 0.95      & 10,866,930       & 78\\
     &      &            &     & 0.95       & 0.95      & 4,831,951,668    & 29,594\\
\hline
     &      &            &     & 1          & 1         & 5,067,570        & 63\\
0.01 & 0.01 &  18,160    & 275 & 0.95       & 1         & 59,233,899       & 422\\
     &      &            &     & 1          & 0.95      & 39,845,600       & 312\\
     &      &            &     & 0.95       & 0.95      & 8,898,295,754    & 54,625\\
\hline
\end{tabular}
\end{center}
\end{table}

The results are listed in Table \ref{table: step3-results}, where \emph{basic operation} refers to comparison, addition, etc.
Here we observe that the dedicated approaches for three subtypes are significantly faster than the one for the general case.
For example, when $ms = mt = 0.01$, approaches for complete match subtype, left-hand side partial match subtype, and right-hand side partial match subtype are 866, 128, and 174 times faster than the general partial match subtype.
Here we focus on the run time instead of the number of basic operations since different operations take different time.
Generally, the speeds of algorithms for three subtypes are 2-3 orders of magnitudes faster than the one for the general case.

  %
  %
  \subsubsection{The performance of different algorithms}\label{subsubsection: performance-algorithms}
\begin{figure}[tb]
    \subfigure[]{
    \begin{minipage}[b]{2.55in}
        \centering
        \includegraphics[width=2.5in]{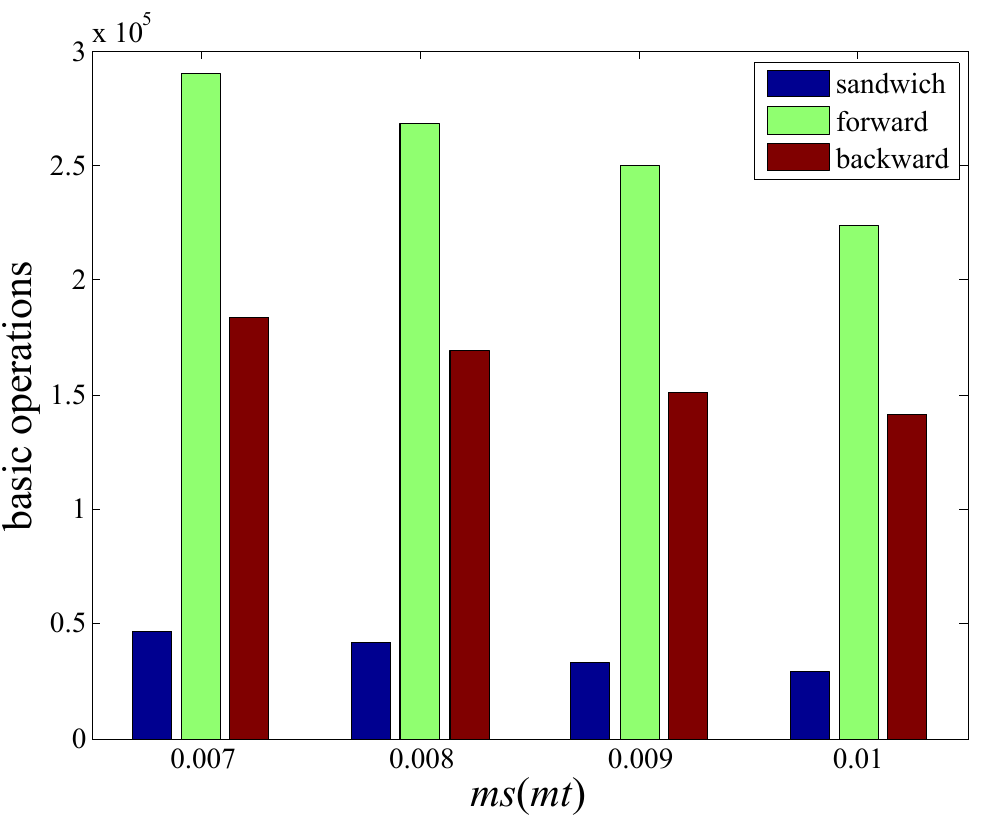}
    \end{minipage}
    }
    \subfigure[]{
    \begin{minipage}[b]{2.55in}
        \centering
        \includegraphics[width=2.5in]{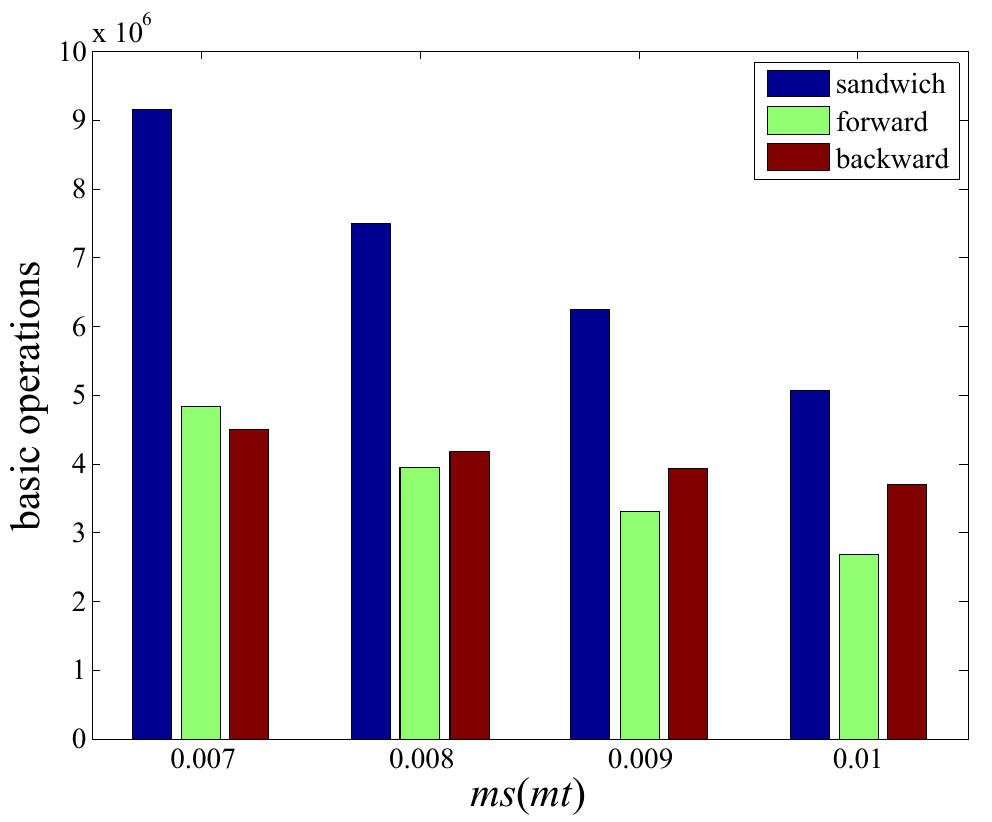}
    \end{minipage}
    }
\caption{Basic operations of three algorithms: (a) MovieLens, (b) Course selection}
\label{figure: algorithmcompare}
\end{figure}

We compare the sandwich algorithm, the forward algorithm and the backward algorithm for the complete match subtype.
Only the number of basic operations are compared, as depicted in Figure \ref{figure: algorithmcompare}.
It is naturally to observe that for the course selection data set, the forward and the backward algorithms are more efficient than the sandwich algorithm.
Moreover, with the decrease of thresholds, the number of operations increases, and the backward algorithm becomes the performs best.
Note that the speed up is not as significant as indicated by Equations (\ref{time-complexity-sandwich}), (\ref{time-complexity-forward}) and (\ref{time-complexity-backward}).
Nor does the backward algorithm outperform the forward algorithm when $|U| < |V|$.
One important reason is that rule checking terminates once certain conditions are met, therefore introducing much uncertainty to the run time.
Consequently, the time complexities are for reference only, and the run time depends more on the characteristics of data.
This might be a common phenomenon for data mining algorithms.

It is very interesting to observe for the MovieLens data set, the sandwich algorithm performs best.
There are at least two reasons.
First, due to threshold settings and data sizes, the MovieLens data set generally requires not too many operations.
Second, the the sandwich algorithm has a simpler mechanism than the other two.
To sum up, the backward algorithm is more scalable.
However for small datasets and large thresholds, the sandwich is more efficient.

  %
  %
  \section{Conclusions and further works}\label{section: conclusion}
In this paper, we have proposed granular association rules to reveal many-to-many relationships in relational databases.
They can be applied to cold-start recommendation \cite{ScheinAI2002Methods,SuXY2009Collaborative} of content-based filtering \cite{BalabanovicM1997Fab,Mooney2000Content}.
Four measures have been defined to evaluate the quality of these rules.
Therefore the new type of rules is semantically richer than existing ones.
We also proposed three algorithms for association rule mining, and compared algorithm efficiency through experimentation.

The following research topics deserve further investigation:
\begin{enumerate}
\item{Different types of data for object description.
In this work we considered only symbolic data for describing objects.
It is necessary to consider numeric data, heterogenous data \cite{HuYuXie08Numerical}, interval valued \cite{Dai12Uncertainty} data and data with missing values \cite{YangWu2006Challenging}.
There are some neighborhood systems concerning distance \cite{HuYuXie08Numerical} or error ranges \cite{MinZhu12Tcsdser} to formalize these data.
Respective approaches (see, e.g., \cite{Dai12Uncertainty,HuYuXie08Numerical,MinZhu12Tcsdser}) can be also employed for these issues.
Moreover, there might be test cost while obtaining data \cite{MinHeQianZhu11Test,MinLiu09AHierarchical}.
Hence we should also consider cost data in certain applications.}

\item{Different granular association rule mining problems.
In the problem definition of this paper, four thresholds are needed as the input.
We may provide other means of parameter setting for non-expert users.
For example, we may mine top-$k$ interesting rules where $k$ is easy to specify.
We may need to remove redundant rules \cite{Pasquier99discoveringfrequent,PeiJ2000Closet} and common sense rules to avoid pattern explosion \cite{VreekenJ2011Krimp}.}

\item{Efficient algorithms to these problems.
As discussed in Section \ref{section: algorithms}, the time complexities of proposed algorithms are rather high.
For datasets with hundreds of thousands of objects, these algorithms may take too much time.
Therefore we need to improve the speed of the algorithms dramatically through taking full advantage of the Apriori property indicated in Section \ref{section: algorithms}.
Rough sets approach to association rule mining \cite{Nguyen99approximatereducts} may be also employed for this purpose.
Moreover, since our algorithms are essentially exhaustive ones, it may be even necessary to design heuristic algorithms for large datasets.
Consequently, we may design heuristic algorithms \cite{VreekenJ2011Krimp} to these problems.}

\item{Theoretical foundations of these problems and algorithms.
The forward and the backward algorithms make use of granule approximation from the viewpoint of rough sets \cite{Pawlak82Rough}, especially the one for two universes \cite{LiZhang08Rough,Liu10Rough,Yao96Two}.
These two algorithms consider only complete match rules; therefore the classical rough set model is employed.
For the general case and two other subtypes, we may need variable precision rough sets \cite{Ziarko93Variable} or decision theoretical rough sets \cite{JiaLiaoTangShang12Minimum,LiZhou2011Risk,LiuD2010Multiple,YaoWong92ADecision}.
There are at least two types of coverings induced by binary relations in this scenario.
The first type of coverings is induced by binary relations.
Given an element in one universe, the binary relation always induces a subset in other.
In this way, from all elements in one universe, a cover of the other universe is induced.
The second type of coverings is induced by granular association rules.
Either side of a rule corresponds to a granule, which describes a covering block.
Covering-based rough sets \cite{LiuZhu08TheAlgebraic,Zhu09RelationshipAmong,ZhuWang03Reduction} are a natural approach for these issues.}

\item{Extension to more than two universes.
From the viewpoints of both theory and application, it is very important to generalize the approach to more than two universes.
Let us consider three universes \texttt{customer}, \texttt{product}, and \texttt{supplier}, and two binary relations \texttt{buys} and \texttt{supplied-by}.
We may chain $n - 1$ granular association rules from $n - 1$ MMERs into one.
For example, we may have the following association rule ``40\% men like at least 30\% kinds of alcohol, 25\% alcohol are supplied by at least 35\% China wine factories."
In this way, mining granular association rules from $n$ universes is decomposed into $n - 1$ subtasks.
We may also multiply the boolean relation matrices to produce a direct relation between the first and the $n$-th universe.
For example, we may multiply \texttt{buys} and \texttt{supplied-by}, and produce a direct relation between \texttt{customer} and \texttt{supplier}.
In this way, mining a granular association rule from $n$ universes is converted into the problem of the paper.
More complex approaches may produce more flexible rules.}
\end{enumerate}

To sum up, granular association rule mining is a challenging problem due to pattern explosion \cite{VreekenJ2011Krimp}.
It may benefit from rough sets, especially variable precision rough sets \cite{Ziarko93Variable} and covering-based rough sets \cite{ZhuWang03Reduction}.
Therefore this work has opened a new research trend concerning granular computing, association rule mining, and rough sets.

  %
  %

  %
  %

\end{document}